\documentclass[prl,twocolumn,superscriptaddress,preprintnumbers,a4paper,amsmath,amssymb,showpacs,floatfix]{revtex4}
\usepackage{graphicx}
\usepackage{bm}

\begin{document}

\newcommand{\btau}{\mbox{\boldmath$\tau$}}
\newcommand{\inA}{\mbox{\AA $^{-1}$}}
\newcommand{\q}{$\delta$}
\newcommand{\qlMn}{$q_{l}^{\rm Mn}$}
\newcommand{\qlDy}{$q_{l}^{\rm Dy}$}
\newcommand{\qlTb}{$q_{l}^{\rm Tb}$}
\newcommand{\qmMn}{$\tau^{\rm Mn}$} 
\newcommand{\qmDy}{$\tau^{\rm Dy}$}
\newcommand{\qmTb}{$\tau^{\rm Tb}$}
\newcommand{\qMn}{$q^{\rm Mn}$}
\newcommand{\qmn}{$\delta_{m}^{\rm Mn}$}
\newcommand{\qdy}{$q^{Dy}$}
\newcommand{\dmn}{$\delta^{\rm Mn}$}
\newcommand{\ddy}{$\delta^{\rm Dy}$}
\newcommand{\bb}{$\mathbf{b}^{*}$}
\newcommand{\qc}{$\delta_{m}=\frac~{1}{4}$}
\newcommand{\qcs}{$\delta_{l}=\frac~{1}{2}$}
\newcommand{\TNMn}{$T_{N}$}
\newcommand{\TNDy}{$T_{N}^{\rm Dy}$}
\newcommand{\TlDy}{$T_{l}^{\rm Dy}$}
\newcommand{\TNTb}{$T_{\rm N}^{\rm Tb}$}
\newcommand{\Tn}{$T_{N}$}
\newcommand{\TN}{$T_{N}$}
\newcommand{\Tc}{$T_{c}$}
\newcommand{\etal}{\textit{et al.}}
\newcommand{\degg}{$^{\circ}$}
\newcommand{\pa}{$\mathbf{P}\|a$}
\newcommand{\pb}{$\mathbf{P}\|b$}
\newcommand{\pc}{$\mathbf{P}\|c$}
\newcommand{\hh}{$\mathbf{H}$}
\newcommand{\Ps}{$P_{s}$}
\newcommand{\Psvect}{$\mathbf{P}_{s}$}
\newcommand{\mc}{$\mu C/m^{2}$}

\title{Coupling of frustrated Ising spins to magnetic cycloid in multiferroic TbMnO$_{3}$}

\author{O.~Prokhnenko}
\affiliation{Hahn-Meitner-Institute, Glienicker Str.~100, D-14109 Berlin, Germany}

\author{R.~Feyerherm}
\affiliation{Hahn-Meitner-Institute, c/o BESSY, D-12489 Berlin, Germany}

\author{M. Mostovoy}
\affiliation{Zernike Institute for Advanced Materials, University of Groningen, 9747 AG Groningen,
Netherlands}

\author{N.~Aliouane}
\affiliation{Hahn-Meitner-Institute, Glienicker Str.~100, D-14109 Berlin, Germany}

\author{E.~Dudzik}
\affiliation{Hahn-Meitner-Institute, c/o BESSY, D-12489 Berlin, Germany}

\author{A.U.B.~Wolter}
\affiliation{Hahn-Meitner-Institute, c/o BESSY, D-12489 Berlin, Germany}

\author{A.~Maljuk}
\affiliation{Hahn-Meitner-Institute, Glienicker Str.~100, D-14109 Berlin, Germany}

\author{D.~N.~Argyriou}
\affiliation{Hahn-Meitner-Institute, Glienicker Str.~100, D-14109 Berlin, Germany}

\date{\today}
\pacs{61.12.Ld, 61.10.-i, 75.30.Kz, 75.47.Lx, 75.80.+q}

\begin{abstract}
We report on diffraction measurements on multiferroic TbMnO$_{3}$ which demonstrate that the Tb- and Mn-magnetic orders are coupled below the ferroelectric transition $T_{\rm FE}= 28$~K. For $T<T_{\rm FE}$  the magnetic propagation vectors (\btau) for Tb and Mn are locked so that \btau $^{\rm Tb}$= \btau $^{\rm Mn}$, while below \TNTb= 7~K we find that \btau $^{\rm Tb}$ and \btau $^{\rm Mn}$ lock-in to rational values of 3/7 \bb\ and 2/7 \bb, respectively, and obey the relation 3\qmTb - \qmMn = 1. We explain this novel matching of wave vectors within the frustrated ANNNI model coupled to a periodic external field produced by the Mn-spin order. The \btau $^{\rm Tb}$= \btau $^{\rm Mn}$ behavior is recovered when Tb magnetization is small, while the \qmTb= 3/7 regime is stabilized at low temperatures by a peculiar arrangement of domain walls in the ordered state of Ising-like Tb spins.
\end{abstract}

\maketitle

While ferroelectricity (FE) and magnetism are chemically incompatible \cite{Hill}, it has recently been shown that inversion and time-reversal symmetry can be broken simultaneously if magnetic spins order in a cycloidal arrangement \cite{kenz,mostovoy,katsura}.  This mechanism is realized in $R$MnO$_{3}$ perovskite manganites ($R=$Tb and Dy) where the magnetic ordering of Mn-spins forms a cycloid within the $bc-$plane.  Accordingly, the spontaneous ferroelectric polarization, \Psvect, is given by $\mathbf{P}_{s}\propto m_{y}m_{z}(\mathbf{e_x}\times\btau)$, where $m_{y}$ and $m_{z}$ are the components of the Mn spins in the cycloid along the $b-$ and $c-$axis respectively, $\mathbf{e_x}$ is a unit vector along the axis of rotation of Mn-spins and {\btau} is the magnetic propagation vector \cite{mostovoy}. 
Recent studies showed that direct and indirect contributions of rare earth spins to electric polarization can also be very important. In DyMnO$_{3}$, for example, the tripling of {\Psvect} is quantitatively accounted by the effective enhancement of $m_{y}$ by the ordering of Dy-spins along the $b-$axis \cite{goto,prok}. Further influence of $R$-ions on the properties of multiferroics are found in the flop of the cycloid into the $ab-$plane (and consequently, \Psvect $\| a$) for the case of the non-magnetic $R=$(Eu,Y) \cite{hemb}. Although the cycloidal ordering of Mn-spins drives multiferroicity, $R$-ions strongly modulate it and thus significantly influence multiferroic properties.

Similar effects are also evident in the $R$Mn$_{2}$O$_{5}$ class of multiferroics despite the fact that here multiferroic behavior may arise from an exchange striction mechanism \cite{chapon,chapon1}. For example, in $R$~=~Tb, \Psvect\ increases in the low temperature incommensurate phase, while it decreases and reaches a negative value for non-magnetic $R$~=~Y \cite{kagomiya,hur}. This difference in behavior results from the influence of Tb-spin ordering upon the magnetic ordering of Mn-spins \cite{chapon}. It is therefore clear that irrespective of the mechanism that drives multiferroic behavior, the magnetic coupling between $R$- and Mn-spins ($J_{{\rm Mn}-R}$) needs to be understood in order to arrive at a detailed and quantitative model of multiferroics and understand ways to improve their performance.

In this Letter we report on diffraction measurements which demonstrate that the Tb- and Mn- magnetic ordering in multiferroic TbMnO$_3$ remain coupled for $T<T_{\rm FE}$. While below $T_{\rm FE}$ it is known that Tb-spins are induced to order with \btau $^{\rm Tb}$= \btau $^{\rm Mn}$ \cite{kenz}, below 7 K we find that \btau $^{\rm Tb}$ and \btau $^{\rm Mn}$ lock-in to rational values of 3/7~\bb\ and 2/7~\bb, respectively, and hold the relationship 3\qmTb - \qmMn = 1.  We explain the novel matching of Tb and Mn wave vectors within the frustrated ANNNI model coupled to a periodic external field produced by the Mn-spin ordering. The \btau $^{\rm Tb}$= \btau $^{\rm Mn}$ behavior is recovered and we show that surprisingly the \qmTb = 3/7 regime is stabilized by an optimal Tb spin-density wave ordering with 6 domain walls, superimposed on the \qmMn = 2/7 Mn-ordering. Our phase diagram not only reproduces the experimentally observed value of \qmTb\ but also explains magnetic  ordering for $R$-spins that is realized for $R=$ Dy and Ho \cite{feyerherm,munoz,brinks}.  

High quality single crystals of TbMnO$_{3}$ were obtained using the floating zone technique and characterized using magnetization measurements with excellent agreement with published data \cite{kimura,kimura3}. Single crystal neutron diffraction measurements were performed on two TbMnO$_{3}$ crystals using $\lambda =$ 2.43 \AA, at the E1 triple-axis spectrometer operated in 2-axis mode and located at the BENSC facility of  the Hahn-Meitner-Institute (HMI). To directly measure the magnetic ordering of Tb-spins and compare it to the ordering of Mn-spins obtained from neutron diffraction we have used  X-ray resonant magnetic scattering (XRMS) at the Tb L$_{2,3}$ edges and 12.4 keV non-resonant X-ray diffraction (XRD) at the 7~T multipole wiggler beamline MAGS, operated by the HMI at the synchrotron source BESSY \cite{dudzik}. For the XRMS measurements, linear polarization analysis was carried out using HOPG(006) and Al(111) crystals, respectively. The $a-$axis was kept parallel to the scattering plane. Additional XRMS measurements with applied magnetic field $H\|a$ and $a$ perpendicular to the scattering plane were performed at the XmaS beamline at the ESRF, Grenoble.

Our single crystal neutron diffraction experiments below $T_N^{\rm Mn}$ largely confirm previous measurements\cite{Quezel, kajimoto, kenz}. We find a set of magnetic satellites in Brillouin zones ($hkl$) with extinction conditions $h+k=even$, $l=odd$ that arise exclusively from the ordering of Mn-spins. Using the convention of Bertaut these are known as A-type reflections \cite{Bertaut}. The magnetic ordering of Mn-spins is incommensurate and characterized by the propagation vector $\mbox{\boldmath$\tau$}^{\rm Mn}$ $\sim$ 0.285 \bb. For $T < T_{\rm FE} = 28$~K, i.e., in the ferroelectic phase, reflections in all G, C, F-type Brillouin zones are observed \footnote{G: $h+k=odd$, $l=odd$; C: $h+k=odd$, $l=even$; F: $h+k=even$, $l=even$.}. These reflection arise from the polarization of the Tb-spins along the $a-$axis \cite{kenz,nadir2}. Below $T_{N}^{\rm Tb}=$ 7~K, a strong increase of intensities of G-, F- and C-type satellites with \btau $^{\rm Mn} = 0.277(2)$~\bb\ is observed and much more intense magnetic satellites appear at positions characterized by a propagation vector \btau $^{\rm Tb} = 0.426(2)$~\bb\ and its odd harmonics. The latter wave vector has been interpreted as to describe a separate ordering of Tb-spins below \TNTb. However, the temperature dependence of $\tau^{\rm Mn}$ reflections suggest that Tb-spins may also continue to order in this regime with \qmMn. 

\begin{figure}[bt!]
\begin{center}
\includegraphics[scale=0.8]{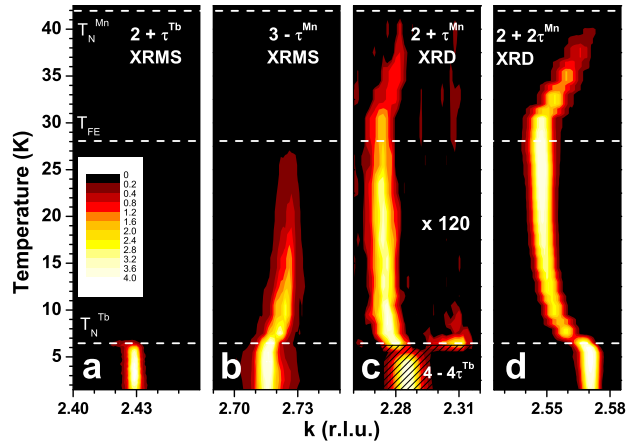}
\caption{(Color online) Temperature dependence of characteristic magnetic and lattice distortion reflections in the form of a color intensity plot measured using XRMS and XRD. In panels (a) and (b) we show $\sigma-\pi '$  data measured from the $\tau^{\rm Tb}$ and $\tau^{\rm Mn}$ peaks in the (0~$k$~2) zone with the X-ray energy tuned to the Tb L$_ 2$ edge. In panels (c) and (d) we shown XRD data of the $\tau^{\rm Mn}$ and 2$\tau^{\rm Mn}$ reflections measured in the (0~$k$~3) and (0~$k$~2) Brillouin zones, respectively. The $\tau^{\rm Mn}$ reflections in panel (c) is $2\times10^{9}$ smaller in intensity than the neighboring $(0 2 3 )$ reflections and represent only magnetic scattering. Below \TNTb~the weak $(0,2+\tau^{\rm Mn},3)$ reflection is superposed by a 100 times stronger reflection from a strucural distortion, $-4\tau^{\rm Tb}$, related to the Tb ordering. The intensity scale in panel (c) is broken close to \TNTb~ and a linear background has been subtracted. The 2$\tau^{\rm Mn}$ in panel (d) is the expected second harmonic structural reflection}. 
\label{fig1}
\end{center}
\end{figure}

In contrast to previous investigations we separate the contribution of Tb spin ordering from that of Mn to the various superlattice reflections by employing XRMS. In Fig.~\ref{fig1}a,b we show contour plots of the temperature dependences of selected reflections. Below $T_{\rm FE}$, a number of relatively weak resonant $(0,k\pm\tau^{\rm Mn},l)$ reflections of G- and C-type  have been observed in the $\sigma-\pi '$ polarization channel, representing the magnetic ordering of Tb with the same propagation vector as the Mn ordering, $\tau^{\rm Tb}_{1}=\tau^{\rm Mn}=0.274(1)$ at 27~K. These reflections are observed below $T_{\rm FE}$ and their intensity increases with concave curvature on cooling -  a behavior typical for an induced moment. With decreasing temperature below \TNTb\, the value of $\tau^{\rm Tb}_{1}$ shifts to a higher value 0.286(1)  but more surprisingly we find no decrease in its intensity in the magnetic channel, a behavior also noted in our neutron diffraction data. Rather we observe a continuous increase suggesting that the magnitude of Tb-spins at this wave vector increases below \TNTb. This effect is accompanied by the appearance of much stronger $(0,k\pm\tau^{\rm Tb}_{2},l)$ reflections, with A- and F-type being dominant, corresponding to a separate ordering of Tb with $\tau^{\rm Tb}_{2}=0.4285(10)$. The observed sequence of magnetic transitions corresponds to Tb magnetic order with $\tau^{\rm Tb}_{1}=$ \qmMn = 0.274 for \TNTb $<T<T_{\rm FE}$ and Tb magnetic order with essentially two wave vectors  $\tau^{\rm Tb}_{2}=$0.4285 and $\tau^{\rm Tb}_{1}=0.286$ below \TNTb. 

To confirm that the shift of $\tau^{\rm Tb}_{1}$=0.274 $\rightarrow$ 0.286 at \TNTb\ (on cooling)  also reflects  the behavior of the Mn sublattice (i.e. the coupling $\tau^{\rm Tb}_{1}=$ \qmMn\ is valid for all $T<T_{\rm FE}$), we employed XRD. In Fig.~\ref{fig1}c,d we show temperature dependent measurements of the non-resonant A-type reflection $(0,2+\tau^{\rm Mn},3)$ (which has only $2\times 10^{-9}$ the intensity of the neighboring (0~2~3) reflection) and the second harmonic reflection $(0,2+2\tau^{\rm Mn},2)$, representing directly the Mn magnetic ordering and the accompanied lattice distortion, respectively. Below \TNTb, both peaks shift simultaneously showing that indeed \qmMn~shifts from 0.274 at \TNTb\ to 0.286 just below it. Having established the behavior of the Tb-ordering, from here on we simplify our notation by taking $\tau^{\rm Tb}( = \tau^{\rm Tb}_{2}$) and \qmMn (= $\tau^{\rm Tb}_{1}$).

We note that below \TNTb\ our x-ray scattering data reveals that to an accuracy of $10^{-3}$ the two wave vectors are locked-in to the rational values \btau $^{\rm Tb}=3/7$~\bb\ and \btau $^{\rm Mn}=2/7$~\bb. These values are slightly larger than those reported by neutron diffraction \cite{kajimoto,kenz}, where the shift of \qmMn~across \TNTb~is much smaller than observed here \footnote{This discrepancy appears to be related to surface strain effects, since the present x-ray scattering measurements probe a thin surface-near region of a few $\mu m$ only and the neutron measurements can be reproduced using 100 keV x-rays (J. Strempfer $\it{et~al.}$, manuscript in preparation). Our basic conclusions are unaffected by this.}. The lock-in of $\tau^{\rm Tb}$ and $\tau^{\rm Mn}$ to values of rational fractions reveals that below \TNTb\ a portion of Tb-spins does not in fact order with yet a separate wave vector, a behavior that would be unusual. Rather, the two wave vectors are related by 3\qmTb~- \qmMn~=~1, implying a single phase Tb magnetic structure below $T_{N}^{\rm Tb}$. This relation does not only hold for the present x-ray scattering data but also for our  neutron diffraction measurements, and others reported \cite{kajimoto,kenz}, to an accuracy of better than 0.002. Therefore, this relation appears to hold irrespective if the values of \qmMn~and \qmTb~are exactly 2/7 and 3/7. The behavior we describe here clearly points to an intimate relation between the Mn and Tb magnetic orderings mediated by $J_{\rm Mn-Tb}$ exchange and suggests that the ordering of Tb spins as well as the ground state wave vectors are determined by this interaction.

The 3\qmTb~- \qmMn~= 1 ordering can be modulated by applying a magnetic field along the $a-$axis. Using XRMS measurements with \mbox{$H\|a$} at 2 K (not shown), we find a first order phase transition at 1 T characterized by 
vanishing of $\tau^{\rm Tb} = 3/7$ reflections and the appearance of new reflections at \btau $^{\rm Tb}$ = 0.361(1)~\bb, and a shift of $\tau^{\rm Mn}$ from 0.285(1) down to 0.276(1). This behavior suggests that now the two wave vectors become related by 2\qmTb+\qmMn=~1.

\begin{figure}[tbp]
\includegraphics[width=8cm]{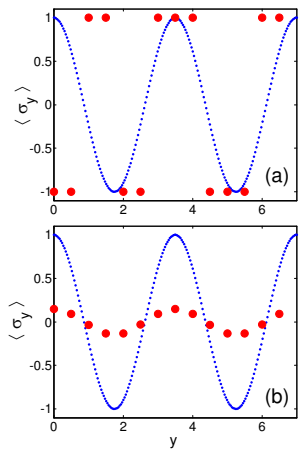}
\caption{(color online) Average values of $R$ spins (red circles) described by the frustrated Ising model in periodically varying external field (dotted blue line) at zero temperature (a) and at high temperatures (b). In the latter case $\tau^{\rm Tb} = \tau^{\rm Mn} = 2/7$, while at $T = 0$, $\tau^{\rm Tb} = 3/7$.}
\label{fig:ordering}
\end{figure}

Phenomenologically, the zero field behavior results from the quartic coupling $g L_{\rm Tb}^3 L_{\rm Mn}$ between the Tb and Mn magnetic order parameters, which favors a relation of the type $3 \tau^{\rm Tb} \pm \tau^{\rm Mn} = n$ ($n$ is an integer number)  between the wave vectors of the two spin-density waves and, in particular, gives the observed relation $3 \tau^{\rm Tb} - \tau^{\rm Mn} = 1$ for TbMnO$_{3}$ at $H$ = 0. In applied magnetic field $H$, the Tb spin-density wave acquires a homogeneous component $\chi H$ and the quartic term effectively leads to the coupling $3 g \chi H L_{\rm Tb}^2 L_{\rm Mn}$ favoring 
$2 \tau^{\rm Tb} \pm \tau^{\rm Mn} = n$, explaining the switch of the relation between the Tb- and Mn- spins observed for \mbox{$H\|a$}.

This phenomenological description applies close to the Tb ordering temperature. At zero temperature Tb-spins are Ising-like \cite{bielen,bouree} and cannot be described by a small number of Fourier harmonics. To understand  the interplay between the $R$ and Mn orderings we consider the one-dimensional ANNNI Ising model,
\begin{equation}
E = \sum_{y} \left[J \sigma_{y} \sigma_{y+1/2}  + J' \sigma_{y} \sigma_{y+1} - \lambda \sigma_{y} \cos(Q y  + \phi)\right],
\label{eq:ANNNI}
\end{equation}
where the Ising variables $\sigma_y = \pm 1$ describe $R$ spins at the sites $y = 0,1/2,1,3/2,2,\ldots$, coupled to the periodically varying `external field' induced by Mn spins, so that $Q = \tau^{\rm Mn}$ and $\lambda \propto J_{\rm Mn-Tb}$. The unit cell of the chain contains two spins (one with an integer and one with an half-integer coordinate) corresponding to two $R$ ions in each $ab-$layer of the unit cell of $R$MnO$_3$.  For ferromagnetic nearest-neighbor coupling, $J < 0$, antiferromagnetic  next-nearest-coupling, $J' > 0$, and $\lambda = 0$ we recover the usual  ANNNI model, which has the FM ground state ($\tau^{\rm Tb} = 0$) for $\kappa = \frac{J'}{|J|}<\frac{1}{2}$ and the  $\uparrow \uparrow \downarrow \downarrow$-state ($\tau^{\rm Tb} = 1/2$), for $\kappa >\frac{1}{2}$. At the frustration point $\kappa = 1/2$ the FM domain wall (DW) energy $E_{DW} = -2(2J'+J)$ is zero and the ground state configuration can have any number of domain walls provided that they do not occupy nearest-neighbor sites \cite{fisher}.

\begin{figure}[tbp]
\includegraphics[width=8cm]{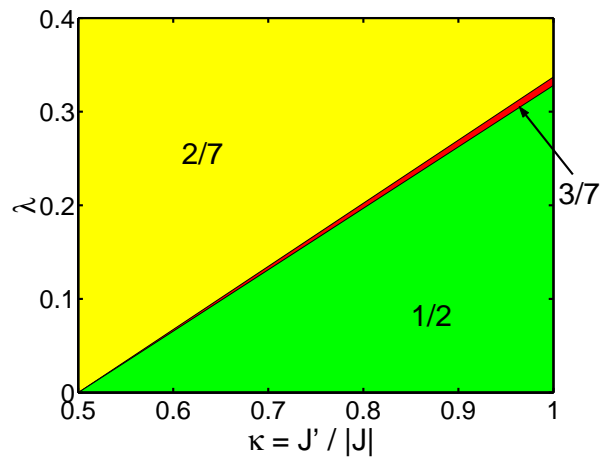}
\caption{(color online) Zero temperature phase diagram of model (\ref{eq:ANNNI}). The phases are numbered by the wave vector $\tau^{\rm Tb}$ of the main spin modulation: $\tau^{\rm Tb} = 2/7$ (yellow), $\tau^{\rm Tb} = 3/7$ (red), $\tau^{\rm Tb} = 1/2$ (green).}
\label{fig:phdiag}
\end{figure}

We now compare energies of the states with $\tau^{\rm Tb}$~=~2/7, 3/7 and 1/2 corresponding to, respectively,  4, 6, and 7 DWs in the interval of 7 unit cells (14 spins). For $\kappa > 1/2$ the DW energy is negative and the $\uparrow \uparrow \downarrow \downarrow$-state having the highest number of DWs has the lowest energy, if only interactions between Tb-spins are considered. The energy of the Tb-Mn interaction in $\uparrow \uparrow \downarrow \downarrow$-state is zero, while for the state with $\tau^{\rm Tb}= \tau^{\rm Mn}$~=~2/7 it is $- 0.642 \lambda$ per spin. Surprisingly,  the state with $\tau^{\rm Tb}$~=~3/7 also gains a substantial energy due to the interaction with Mn spins, which for the best possible arrangement of 6 DWs, shown in Fig.~\ref{fig:ordering}a, is $-0.218 \lambda$ per spin.

The resulting zero temperature phase diagram is shown in Fig.~\ref{fig:phdiag}. For the weak $R$-Mn coupling regime, the  $\uparrow \uparrow \downarrow \downarrow$-state has the lowest energy and the wave vectors of $R$ and Mn orderings are decoupled as is the case for $R$ = Dy \cite{feyerherm}. For large $\lambda$, $\tau^{\rm R}= \tau^{\rm Mn}$ down to zero temperature as found for $R$ = Ho \cite{munoz,brinks}. In the narrow sector $0.656 < \frac{\lambda}{2 J' + J} < 0.674$, Ising spins order in the ground state with $\tau^{\rm Tb}$~=~3/7, as observed in TbMnO$_3$. This is a rather unexpected result since the 3/7-state is stabilized by the modulation of Mn-spins with a different wave vector. Numerical mean field studies of the three-dimensional version of model (\ref{eq:ANNNI}) for parameters favoring the 3/7-state show that at high temperatures the wave vector of the Tb modulation equals that of Mn-spins (see Fig.~\ref{fig:ordering}b), while below \TNTb\ the intensity of the $\tau^{\rm Tb}=$~3/7 peak is several times higher than that of the  $\tau^{\rm Tb}=$~2/7 peak, in agreement with our diffraction data. The small extent of the 3/7-region in the phase diagram of our model suggests that other mechanisms, e.g. longer range interactions between $R$-spins and the deformation of the Mn spin-density wave in response to the ordering Tb-spins, may contribute to stabilization of this phase.

Our results uncovered a strong coupling between Tb- and Mn-spin sublattices in multiferroic TbMnO$_{3}$ that is allowed because of the FE nature of this compound at low temperatures. Indeed, Mn magnetic ordering alternating in the $c$-direction should produce zero exchange fields at the $R$-ion as noted for HoMnO$_{3}$ in  Ref. \cite{munoz}.  We believe that this argument is true above the FE transition in $R$ = Tb, Dy and Ho, where consequently no induced moment is found. In turn, the observation of induced ordering below the FE transition can be considered as strong evidence of symmetry breaking that would allow $R$-moments to be induced as the spontaneous electric polarization occurs. 
 
In summary, our work demonstrates that the $J_{\rm Mn-R}$ interaction is a critical ingredient in our understanding of the magnetism and physical properties of multiferroic manganites with magnetic $R$-ions, highlighting the relevance of our model. In the temperature range $T_{N}^{\rm R}<T<T_{\rm FE}$ the coupling of $R$- and Mn-spin order is reflected in $\tau^{\rm R}=\tau^{\rm Mn}$, a behavior that is now confirmed for $R=$Tb, Dy and Ho \cite{kenz, feyerherm,munoz}. For $T<T_{N}^{\rm R}$, $R=$~Ho represent a strong coupling regime as  \qmMn= $\tau^{\rm Ho}$ is preserved  to the lowest temperatures \cite{munoz,brinks}. Our model reproduces the behavior for $R$=Dy for a weak coupling regime as we recover the commensurate $\tau^{\rm Dy}$=1/2 ordering \cite{feyerherm,prok}. However, the case of $R=$ Tb is more interesting as it represents an intermediate coupling regime with complex behavior. In particular, below $T_{N}^{\rm Tb}$, when $J_{\rm Tb-Tb}$  becomes important, we find  that the Tb- and Mn- orderings remain coupled through the matching of their wave vectors, $3 \tau^{\rm Tb} - \tau^{\rm Mn} = 1$.  The adjustment of Ising-like Tb spins to periodic Mn ordering results in a rather complex shape of the Tb spin-density wave at low temperatures. 

Finally, the coupling between $R$- and Mn-sublattices in $R$MnO$_3$ with magnetic $R$ = Tb, Dy and Ho may have some implications for magnetically-induced ferroelectricity in these multiferroics. Referring to Eu$_{1-x}$Y$_{x}$MnO$_{3}$ with nonmagnetic $R$ and Mn-cycloid within the $ab$-plane \cite{hemb}, the stabilization of the $bc$-cycloid for $R$ = Tb and Dy may be caused by the energy gain due to the Tb-Mn exchange in the FE state, which allows non-zero Mn exchange fields at $R$ sites. The counterintuitive magnetic field behavior for $R$ = Tb and Dy, e.g. the cycloid flop from the $bc$- to $ab$-plane in \mbox{$H\|a$}, can then result from reduction of the $R$-Mn interaction when $R$-spins become aligned ferromagnetically by the magnetic field.

The authors have benefited from discussions with D. Khomskii and K. Prokes. We thank Vadim Sikolenko and Roland Schedler from HMI, and Simon Brown and Laurence Bouchenoire from ESRF for assistance during the experimental work. M.M. thanks HMI for hospitality.

\end{document}